\documentclass[12pt,preprint]{aastex}

\newcommand{\etal}{$et \; al.$~}

\def\arcsec{\hbox{$^{\prime\prime \, }$}}
\def\micron{\hbox{$\mu$m}}
\def\lea{\mathrel{<\kern-1.0em\lower0.9ex\hbox{$\sim$}}}
\def\gea{\mathrel{>\kern-1.0em\lower0.9ex\hbox{$\sim$}}}

\begin{document}

\title{A V-Band Survey for Variable Galactic Nuclei in the Hubble Deep Field}
\author{Vicki L. Sarajedini}
\affil{University of Florida, Gainesville, FL 32611}
\email{vicki@astro.ufl.edu}
\author{Ronald L. Gilliland}
\affil{Space Telescope Science Institute, 3700 San Martin
Drive, Baltimore, MD 21218}
\email{gillil@stsci.edu}
\and
\author{Christina Kasm}
\affil{University of Florida, Gainesville, FL 32611}
\email{ckasm@astro.ufl.edu}

\begin{abstract}

We present the results of a 2-epoch variability survey in 
the Hubble Deep Field with the goal of
investigating the population of AGN to z$\simeq$1.
The high resolution and stability of HST allows 
accurate photometry to be obtained within subarcsecond apertures,
resulting in the ability 
to probe much lower AGN/host galaxy luminosity ratios than can be done
from the ground. 
The primary data sets analyzed for galactic variability are the original
HDF observations obtained in December 1995 and a second 
V-band (F606W) image obtained almost exactly 5 years later in 2000.
We find evidence for nuclear variability in 16 of 217 galaxies brighter
than V$_{nuc}$$=$27.5.  Correcting for incompleteness and spurious
detections, variable nuclei make up $\sim$8\% of the surveyed galaxies.
These sources have a redshift range of  
0.09$<$z$<$1.8 and cover the full range of galaxy nuclear V--I colors.
Seven of our variable sources are coincident
with X-ray sources detected in the 2Ms Chandra survey; six from
the main catalog and one from the supplementary catalog. 
We find that 44\% of the variable nuclei are associated 
with mid-IR detections at 15$\micron$ and 31\% are detected at 1.4GHz.
Optical spectra are available for 13 of the 16 variables.  One is
a broad-line AGN and 2 others show weak evidence of
type 2 AGNs.  With the assumption that these variables are all 
active nuclei, we estimate the AGN LF  
at 0.4$<$z$<$1.1 extending to M$_B$$\simeq$-15.  We find evidence
for an increase in the number density of faint AGN when comparing
to the local Seyfert luminosity function.  The LF for optically
varying nuclei appears to rise in number density 
with no evidence of turning over at these faint magnitudes.

\end{abstract}


\section{INTRODUCTION}

An important goal of modern cosmology 
is to understand the evolution of active galaxies and their
relationship to normal galaxy evolution.
A key piece of information needed to further this understanding is an
accurate knowledge of the luminosity function (LF) for AGN over a wide range of
absolute magnitudes and redshifts.
The AGN LF is populated by
quasars at the bright end (e.g. Hartwick \& Schade 1990) and
Seyfert galaxy nuclei, considered to be their intrinsically fainter
counterparts, at the low luminosity end (Cheng \etal 1985;
Huchra \& Burg 1992; Maiolino \& Rieke 1995; Koehler \etal 1997;
Ulvestad \& Ho 2001; Londish \etal 2002).

Understanding how the faint end of the AGN LF evolves
is of particular importance for determining the frequency and
total space density of AGNs at earlier epochs.  This has
obvious implications for determining their total contribution
to the X-ray, IR and UV backgrounds.  Large numbers of low-luminosity
AGN have been proposed to explain the ionization of the intergalactic
medium at high redshift (Steidel \& Sargent 1989), although
recent observational and theoretical results differ on the amount of their
true contribution (Barger 2003; Schirber \& Bullock 2003).
Additionally, the faint
end is an important constraint on evolutionary models for
the AGN LF such as pure luminosity and luminosity-dependent
density evolution (e.g. Boyle \etal 2000).

One method to identify AGN in imaging surveys is via their variable nature.
QSOs and Seyfert galaxies have long been known as variable
objects, with significant optical flux changes occurring on
timescales of months
to several years.  Variability has been a successful
method for identifying AGNs, primarily QSOs
(Hawkins 1986; Koo \etal 1986;  Hook \etal 1994).
A survey for variable sources in
SA57 revealed many optically extended, Seyfert-like galaxies
(Bershady \etal 1998) which generally had higher variability
amplitudes than the more luminous QSOs.  This result suggests that
variability is a good technique for selecting
intrinsically faint QSOs and Seyfert nuclei.

While variability and other selection techniques
(spectroscopy, color selection, etc.) have been successful in finding
QSOs out to high redshifts, the
intrinsically fainter AGN (i.e. Seyfert galaxies) 
have been difficult to detect beyond the
local universe.  Ground-based variability surveys,
like spectroscopic surveys, are quickly limited to AGN which dominate
the host galaxy light at higher redshift.  Recent X-ray surveys with
Chandra and XMM-Newton have revealed hundreds of X-ray sources
at z$\simeq$1.  Many of these are confirmed
AGN but still many more are ambiguous in nature (e.g Barger \etal 2002).
In summary, the population of moderate-to-high redshift, 
low-luminosity AGN has
not been well surveyed and the evolution of the faint end of the
AGN LF is still poorly known.
 
In this study, we have searched for optical nuclear variability
in HDF\footnote{HDF refers to
the Hubble Deep Field North} galaxies to identify the population
of active galaxies out to z$\sim$1.  The primary data sets 
analyzed for galactic variability are the original
HDF observations (Williams \etal 1996) obtained in December 1995 and a second
V-band (F606W) image obtained almost exactly 5 years later in December of 2000.
The main advantage of HST over ground-based
surveys is the ability to do accurate photometry
within smaller apertures than can be done
from the ground thus allowing us to probe much lower
AGN/host galaxy luminosity ratios.

We discuss the details of the image processing and photometry
in Section 2, followed by a discussion of the selected varying galactic nuclei
in Sections 3 and 4.  In Section 5, we cross-reference our sources with various 
multi-wavelength surveys.  
A discussion of the optical spectra for the varying sources is
presented in Section 6.  In Section 7, we present the luminosity function for
variable nuclei.  We close with a summary of our conclusions in
Section 8.

\section{PHOTOMETRY}

\subsection{Image Processing}

The original epoch HDF data in F606W consisted of 108 individual
exposures with a total integration time of 113,050 seconds.  These data
were obtained with a total of 9 unique dithers covering both sub-pixel
offsets and multiple pixels on a scale of about $\pm$ 10 WF pixels.
The primary data processing technique was detailed in Gilliland, Nugent, and
Phillips (1999) for creation of over-sampled, by a factor of four,
summed images with resulting noise and resolution properties competitive
with, and sometimes slightly better than those reported in
Williams \etal (1996).  The mean epoch of the original F606W
data was MJD 50076.9.  Exposure times used for the original epoch ranged
from 350 to 1500 seconds with a mean sky rate of 0.0758 e-/s.

The second epoch HDF data in F606W was taken with GO-8389 (PI Rodrigo Ibata)
with a science goal distinct from that here, but one that also required
precise differential measurements across the two epochs.
These data consisted of 70 individual exposures each of 1300 seconds
for a total exposure time of 91,000 seconds.  The goal was to have the
new data obtained at the same pointing as the original HDF, and at the
same orientation, and with dithering over the same spatial scale, but
with an improved level of sampling using 35 unique dithers.  This goal
was quite well met, thus resulting in quite similar data in the two
F606W epochs.  The mean epoch of the second F606W data was MJD 51882.9
implying a delta of 4.94 years.  The mean sky rate for the newer epoch
was 0.0958 e-/s.  Processing of the second epoch data was identical to
that of the first, using the same software as described in Gilliland,
Nugent and Phillips (1999).  To facilitate simple, robust analysis of
data from the two epochs the registrations of the second epoch were
evaluated using a grid of galaxies from the original epoch (stars could
not be used since these show unique proper motions over five years);
this results in second epoch over-sampled combined images that are
accurately aligned to the first epoch with {\em maximum} errors of
$\sim$0.05 pixels.  Therefore, comparative analysis for change across 
epochs can be made
using precisely the same pixels, mapping back to common physical domains.
For both epochs, each individual frame was sky subtracted before combining,
cosmic rays were eliminated (effected pixels assigned zero weight in the
combination) and hot pixels subtracted.

The extreme stability provided by {\em HST} is an essential factor
in being able to probe to unique sensitivity levels with these extensive
observations.  Limiting factors, to be discussed further below, arise
from (1) a time-varying Charge Transfer Efficiency (CTE) that must
be empirically corrected for, and (2) limited sub-pixel dithering in the
first HDF epoch that results in minor mismatches noticeable for the
objects with tight cores at high signal to noise.

\subsection{Source Selection and Aperture Size}

We created a catalog of HDF objects using
DAOFIND in IRAF\footnote{
IRAF is distributed by the National Optical Astronomy Observatories,
which are operated by the Association of Universities for Research
in Astronomy, Inc., under cooperative agreement with the National
Science Foundation.} to ensure 
a systematic selection of all sources in the V-band images.
All objects above the 8$\sigma$ detection threshold 
were identified in the 1995 images (the deeper of the two)
using a kernel size of 1.8$\times$ the semi-major axis of the
Gaussian convolution kernel (1.3 pixels) to allow
more sensitivity to extended objects.  The catalog was then
edited to remove a minority of objects that did not appear to be centered
on galaxies (i.e. multiple detections within single galaxies). 
For very amorphous galaxies with no clear center,
all bright ``knots" appearing within the source were 
retained in the catalog.  

For our photometry, we chose a small aperture consistent 
with the FWHM of stellar
sources in the field.  Stars have FWHM$\simeq$5.5 pixels in the over-sampled 
images corresponding
to $\sim$0.14$\arcsec$.  We chose an aperture size of 6 pixels in diameter
to be most sensitive to photometric changes in an unresolved nucleus.
This approach differs from that taken in 
Sarajedini, Gilliland \& Phillips (2000; hereafter SGP), 
where the aperture size
was scaled to the size of the galaxy.  Our goal here is to
provide better sensitivity to variable active nuclei with
less dilution from the underlying, non-varying host galaxy light. 

Aperture photometry for all sources was performed on the direct
images using the PHOT algorithm in IRAF.  Because the second epoch
image was aligned with the first, exactly the same pixels were used
in both epochs.  During image processing, the background was
subtracted and the images were scaled to an exposure time of 6000s.
This allowed the sky value to be held constant at zero
for both epochs and all WF CCDs.  The resulting nuclear magnitudes for
each galaxy in the HDF were then adjusted by a zero-point
offset to bring them into agreement with the integrated, isophotal 
galaxy V-band AB magnitudes of Williams \etal (1996).   

\subsection{Charge Transfer Efficiency Losses And Photometric Errors}

After performing aperture
photometry, nuclear variability
can then be determined by simply obtaining the difference
in magnitude for each source over the two epochs.
Figure 1 shows the magnitude difference
between the 1995 and 2000 images as a function of magnitude within an
aperture of r$=$3 pixels, the equivalent of 0.15$\arcsec$ 
in diameter.  The solid line represents no difference in magnitude
between the two epochs.

The main culprit producing the obvious offset from the solid line
is charge transfer efficiency (CTE) losses.  As in SGP, we find a clear 
correlation of the photometric
differences with position on the CCD and magnitude.  
This is the well known CTE effect (Whitmore \etal 1999; 
Biretta \etal 2001) which causes targets far from the CCD 
readout amplifier to appear fainter than similar targets near the amplifier.
For this reason, the effect is usually more significant in the Y-direction,
(the direction of the readout down the CCD) than the X-direction (the 
direction of the readout of the shift register at the bottom of the CCD). 
Although the 1995--1997 I-band study did not show significant 
correlations in the
X-direction, the longer time interval sampled here
does show a significant correlation in the X-direction
for all 3 WF CCDs.  Additionally, we note that the
slope of the relation is generally steeper in the Y-direction
than had been determined in SGP.
This is expected due to the longer 5 year time interval
between epochs in this study.

We first attempted to fit the relationship between object position,
magnitude difference and magnitude globally for all 3 WF CCDs
as done in SGP.  However, we found that this approach was
not sufficient to equally remove the CTE effect from all WF CCDs.
Differences in the slope of the relation between
the CCDs were significant and could easily be determined for
each CCD individually.   We adopted a two-step approach to determine
this relationship.  First, we fit a linear surface to the X and Y object
positions and the magnitude differences using SURFIT in IRAF for
each CCD.  While the slope of the fit was found to be significant with
both the X and Y positions, the value of the slope was steeper with Y position
in all 3 WF CCDs, consistent with what is expected for CTE losses.  
The slope of the relation between Y position and
magnitude difference was found to be -2.6$\times$10$^{-5}$,
-1.3$\times$10$^{-5}$ and -1.46$\times$10$^{-5}$ mag/pixel for WF2, WF3 and
WF4 respectively.  The slope of the relation with X position, while 
significant, was found to be about an order of magnitude lower in
all 3 WF CCDs.
These fits were used to correct the magnitude difference values 
for position-dependent offsets.
Secondly we fit the dependence of the magnitude difference
with nuclear magnitude.  
A 3rd order fit was required to properly model the dependence in all
3 WF CCDs.
After applying this correction to the data in Figure 1, the resulting
photometric differences are shown in Figure 2 as a function of 
nuclear magnitude.  

Finally, we determine the 
expected noise level for non-variable
sources.  As in SGP, we divided the full set of individual exposures
into two sets composed simply of the odd-numbered frames and the
even-numbered frames.  Each set represents an ``epoch" without any
real time difference.  These images effectively carry through the
effects of object and sky Poisson noise, readout noise, and possible
errors in the adopted sky zero points.

The magnitude of each galaxy was then measured in the odd and even
data sets and difference in magnitude vs. the average magnitude
within r$=$3 pixel apertures is shown in Figure 3a.  Figure 3b is
the RMS of the intensity differences shown in a) within unity
magnitude intervals.  The solid line represents a quadratic fit to the
points which is the adopted 1$\sigma$ galaxy photometric error
as a function of magnitude.  

The precision of the photometry must be independently limited at the
bright end of the distribution, since systematic effects not captured
by the odd-even data set comparisons become limiting factors (e.g.
the sometimes poor distribution of sub-pixel phase-space dithers in
the first epoch data).  In SGP, we adopted a floor of 1.2\%.  Here we
adopt a more conservative value of 1.5\% consistent with the use of 
much smaller apertures on average for bright sources in the current
study than in SGP.  As should be apparent, this is a reasonable limit
for photometric precision to adopt, but in detail remains somewhat
arbitrary.

\section{SELECTION OF VARIABLES AND SURVEY COMPLETENESS}

Figure 4a is the absolute value, CTE-corrected
magnitude differences for sources in the HDF.  
The solid line represents the
3$\sigma$ limit for variability significance (3$\times$ the
RMS indicated by the solid line in Figure 3b).  Objects above this
limit are selected as significant variables and are indicated with 
open hexagons.  Figure 4b is the
result of normalizing the magnitude difference by this
solid line.  Here the Y-axis indicates the level of significance
of each source in units of $\sigma$.  
The X-axis extends at the faint end to V$_{nuc}$$=$29.0,
which is the estimated photometric completeness magnitude limit for 
galaxy nuclei in this survey.  Beyond this limit, the number counts
for galaxies in the HDF begin to decrease.  

We find sixteen galaxies whose nuclei
have undergone at least a 3$\sigma$ variation 
over the 5 year time interval.
These sources are listed in Table 1 with
columns as follows: (1) Williams \etal (1996) ID, (2) \& (3) 
RA \& DEC (J2000), (4)  Redshift from the literature,
(5) Spectral type based on photometry from Fern\'{a}ndez-Soto, Lanzetta
and Yahil (1999, hereafter: FLY),
(6) V$_{nuc}$ internal to r=3 pixel aperture,
(7) Magnitude difference between 1995 and 2000; positive implies brighter
in 1995, (8) Significance of change obtained when normalized by the
expected error as a function of magnitude, (8)
Bulge-to-Total (B/T) 2-dimensional model fits for the I(F814W) images from 
Marleau \& Simard (1998).
The last several columns relate to the source detection at other wavelengths 
discussed in Section 5. 

There are two distinct completeness issues that effect our survey.
The first is related to the incomplete time-sampling of the variable sources
we wish to detect. 
Most variability surveys employ a method where the survey field is
imaged several times over many years (e.g. Trevese \etal 1994;
Hawkins 2002).
Depending on the quality of the data, these surveys have shown that
virtually all
known AGN will be found to vary if observed periodically over
several years.  Our study is limited by
the fact that we have only two epochs with which to determine variability
and therefore sample just two points on the lightcurve of a varying source.
Because of this, we will be incomplete in our census of AGN since
some varying sources could lie at magnitudes close to their
original magnitude measured 5 years earlier and would thus go
undetected in our survey.

We estimate our incompleteness due to undersampling of the lightcurve
by using variability data for
AGN obtained over many years.  We 
randomly select points along the lightcurve 
separated by 5 years, the time interval sampled by our HDF
images.  Because long term variability surveys for low-luminosity AGN
have not yet been published, we have conducted this test with
two sets of QSO lightcurve data;  a sample of PG quasars from
Giveon \etal (1999) and a sample of SA57 quasars from
Trevese \etal (1994).  Giveon \etal
monitored 42 quasars over 7 years with a typical sampling interval
of 40 days while Trevese \etal monitored 64 over 15 years about once
a year.  Sampling these lightcurves every 5 years and assuming
photometric errors typical of our HDF images, we can estimate the
probability that these sources would be detected in our survey.
We find that $\sim$75--80\% would be detected as
varying by at least 3$\sigma$ with only 2 observations separated by 5 years.
Therefore, the results of our analysis are likely to underrepresent
the true number of variable nuclei by $\sim$20--25\%.

The second incompleteness issue results from the use of a fixed
aperture to detect nuclear variability.  We have chosen a small, fixed
aperture to minimize dilution of the AGN light from the underlying
host galaxy to be more sensitive to AGN varying within bright hosts. 
We set this aperture to the size of an unresolved
point source in the HST images (0.15$\arcsec$ in diameter).  
The fixed aperture will include the same amount of light from
an unresolved source regardless of its redshift.  However, the
aperture will contain a larger fraction of the underlying, resolved
galaxy for higher redshift objects than for low-z sources.  Because
of this, the dilution of the nuclear light by the galaxy increases as
a function of redshift and, consequently, as a function of nuclear magnitude.
If we assume a disk-dominated, r$_e$=1 kpc galaxy containing an AGN that
is 10\% of the host galaxy light and varying by $\Delta$m=0.3 magnitudes,
the measured $\Delta$m within our fixed aperture would be 0.15 magnitudes
at z=0.2 but only 0.06 magnitudes at z=1.  The observed total nuclear
magnitude would drop by $\sim$3 magnitudes over this interval.
The increasing dilution with redshift is compounded
by the increasing photometric errors with magnitude.  
For an AGN that has varied by $\Delta$m=0.3--0.2 mags, consistent with
typical structure function values for AGN (Trevese \etal 1994),
and reasonable estimates for the galaxy magnitude, morphology, physical size
and true AGN fraction, we estimate that incompleteness due to significant
AGN dilution begins to effect our survey at magnitudes between 
V$_{nuc}$$\simeq$26.5 to V$_{nuc}$$\simeq$27.5.  

As can be seen in Figure 4, all of the variables lie at magnitudes
greater than V$_{nuc}$=27.3.  While this may partially be due to
a real astrophysical effect (see discussion in Section 4), this
limiting magnitude is within the range of our incompleteness
estimate above.  Therefore, although we survey galaxy nuclei
to V$_{nuc}$$\simeq$29, our limit
for detecting {\it variable AGN} at a uniformly varying level within 
galaxy nuclei is probably closer
to V$_{nuc}$$\simeq$27.5, with decreasing completeness fainter
than V$_{nuc}$$\simeq$26.5.

The 16 variable nuclei detected in
our survey represent 2.2\% of the galaxies brighter than V$_{nuc}$$=$29.0
or 7.4\% of the galaxies brighter than V$_{nuc}$$=$27.5.
To emphasize the significance of the variable sources,
Figure 5 is a histogram of the sigma distribution for galaxies
in the HDF.
The X-axis is the absolute value of the normalized magnitude difference
($\sigma$)
and the Y-axis is the natural logarithm of the number of sources
in 0.25$\sigma$ bins.  Errorbars represent the poisson statistical errors.
Filled circles show the histogram for all 719 nuclei brighter
than V$_{nuc}$$=$29.0 and open squares represent the 217 nuclei
brighter than V$_{nuc}$=27.5.
The curved lines are gaussian fits to the data within 2.5$\sigma$
(solid is the fit to data brighter than V$_{nuc}$$=$29.0 and
dashed is the fit to data brighter than 27.5).
In both distributions,
the data are well fit by gaussians out to $\sim$3$\sigma$ and show
a ``tail'' of significant variables extending to higher $\sigma$ values.
In a normal distribution, we would expect a total of $\sim$5 sources to be
greater than 3$\sigma$ brighter than V$_{nuc}$$=$29.0 with $\sim$1.5
expected to lie brighter than V$_{nuc}$$=$27.5.

Based on Gaussian statistics, we estimate that 
$\sim$1--2 of our variables (all detected brighter
than V$_{nuc}$$=$27.5) are spurious, implying that the true number of variables
is $\sim$14.  If we also correct by 20--25\% for
incompleteness due to lightcurve sampling, the incompleteness corrected
number of variables in the HDF is $\sim$18 representing 8.3\% of the
nuclei brighter than V$_{nuc}$$=$27.5. 

One of the 16 variables detected in this survey, {\it 2-251.0}, 
was also identified
as variable in the I-band variability survey of the HDF (SGP).
Eight variables were detected in that survey with significant ($\gea$3$\sigma$)
magnitude changes over
the 2 year time interval (1995--1997).  Object {\it 2-251.0} is by far the brightest
of these sources, having a nuclear V magnitude $\sim$13$\times$ brighter
than the next brightest source.  Of the 7 I-band variables not detected
in the present survey, 5 are fainter than V$_{nuc}$=28.0, placing them
below the expected sensitivity limits for this study.  The other two
I-band variables, {\it 3-266.0} and {\it 3-404.0}, have V$_{nuc}$$\simeq$27.0.
In the present study, object {\it 3-266.0} changed by 0.025 mag (1.1$\sigma$
significance) and object {\it 3-404.0} changed by 0.052 mag (2.0$\sigma$
significance).  The latter source is a marginally significant detection and
would lie above the 3$\sigma$ limit if it had been brighter than $\sim$26.5
in the V-band.  Object {\it 3-266.0}, at just over 1$\sigma$ significance,
may be variable but at lower significance due to the 2-epoch statistical
incompleteness effect discussed above.

\section{PROPERTIES OF THE SELECTED VARIABLE NUCLEI AND HOST GALAXIES}

Figure 6 contains the 1995 V-band images for galaxies with 
variable nuclei.  The postage stamp images are 3$\arcsec$ on a side
and are scaled to the same maximum pixel value.
It can be seen from this figure and Table 1 
that the selected galaxies cover a range of morphologies, 
magnitudes and colors.  Their
redshifts range from 0.09 to 1.8 with a median $<$z$>$$\simeq$0.66.
Many of the galaxies have significant bulge components based
to the 2-dimensional model fits.  We note, however, that several
sources with B/T$=$0 also have high chi-square values ($>$1.25)
which in some cases appears to be due to a poorly fit bulge component.
In addition to the morphological B/T parameter, we also list the
spectral type for the source based upon the photometry of FLY to
determine photometric redshifts for galaxies in the HDF.  The
selected galaxies appear to be evenly distributed over all
spectral classes with no particular preference towards early or
late-type galaxies.  

Figure 7 is a color magnitude diagram of the galaxy nuclei for
sources in the HDF.  The variable nuclei are
indicated with larger solid triangles.  The variables
cover the full range of V--I colors and do not appear to be
preferentially bluer or redder than the non-varying nuclei of
galaxies in the HDF.

All of the variables lie at the
bright end of the magnitude distribution.
As previously discussed, incompleteness due to increasing AGN 
light dilution as a function of redshift is likely to contribute 
to this observation.
However, an astrophysical effect may be at least partly responsible.
AGN are generally found in brighter host galaxies than those
that do not harbor AGN (Huchra \& Burg 1992; Ho \etal 1997; Hamilton \etal
2002).  If these sources
are indeed AGN, we might expect to find them in the
brighter galaxies.  An additional factor is the inclusion of the AGN
light in the nuclear magnitude.  We can minimize the above mentioned 
incompleteness effect by considering only those sources brighter 
than V$_{nuc}$=26.5, where incompleteness is not expected to be significant.
We find that 23\% (5/22) of
nuclei are variable at 21.8$<$V$_{nuc}$$<$25.5 while only 15\% (6/39) are 
variable at 25.5$<$V$_{nuc}$$<$26.5.  While the small numbers do not provide
significance, this trend is consistent with the previous studies concerning
AGN host galaxies.  

\section{COMPARISON WITH MULTI-WAVELENGTH HDF SURVEYS}

The HDF has been surveyed with a variety of other telescopes
and instruments providing a wonderful resource of multi-wavelength
information for this particular region of the sky.
We have cross-referenced our list of variable nuclei
with published catalogs of X-ray, radio and far-infrared
sources as well as other photometric
and spectroscopic surveys used to search for nuclear activity.

AGN have long been recognized as X-ray sources, being 
associated with a wide range of activity
levels from the brightest quasars to low-luminosity Seyferts.
Even highly obscured AGN can often be detected at X-ray wavelengths.
Some of the light obscured by dust at other wavelengths will
be reprocessed and emitted in the mid-IR.   For these reasons,
the X-ray and mid-IR regimes are excellent wavelengths in which to
detect and study AGN. 

Recent deep surveys with the Chandra X-ray Observatory have
resolved many X-ray sources in the HDF and  
provide information about the nature of the X-ray radiation.
The 2Ms Chandra X-ray survey (Alexander \etal 2003) has detected 20 
sources in the original
HDF and an additional 2 sources of lower significance which are listed in 
their supplementary catalog.  Of these 22 detections, 18 are included
in our variability study.  Two were not included because their optical 
fluxes were too faint to be included in our survey and an 
additional pair fell too close to the edge of 
the CCD to obtain accurate nuclear photometry in both epochs.
Of these 18 sources, we find seven
that match (within 1.2$\arcsec$) the positions of our variable nuclei
(see columns 10 and 11 of Table 1).  The filled symbols
in Figure 4 represent the 18 X-ray sources included in our survey and
show the level of variability significance for each source.

In addition, we have found several variables
coincident with source positions from the ISOCAM 15$\micron$ survey of
the HDF (Aussel \etal 1999) and the 1.4GHz radio survey (Richards
\etal 1999).  These are listed in columns 12 through 14 of Table 1. 
Below, we discuss the variable galaxy nuclei that overlap with one or more
of these multi-wavelength surveys. 

{\it 2-251.0:}  
This z=0.960 early-type spiral galaxy is one of only two broad-line AGN in
the HDF (see Section 6) and is a 
bright X-ray source detected with Chandra in the soft, hard and 
ultrahard bands (Hornscheiemer \etal 2000; Brandt \etal 2001a -- hereafter
H00, B01a) with a photon index of 0.67.  We detect
it as an optical variable at 6.5$\sigma$ significance.  
As previously mentioned, it is the only
object that was also detected in our I-band survey 
of the HDF (SGP).  
Spectroscopic detection of broad MgII (Phillips \etal 1997), as well
as 1.4GHz radio (Richards \etal 1999a) and ISOCAM (Aussel \etal 
1999) detections for this source all
corroborate the AGN nature of this galaxy.  Spectral fitting of
the X-ray data has revealed a large intrinsic column density which
appears to be related to the AGN (B01a).
 
{\it 3-355.0:} 
This elliptical galaxy at z=0.474 is a
3.2$\sigma$ variable.  The original 166ks Chandra exposure 
did not detect this source in the hard band, but B01a report 
hard X-ray counts from the deeper 479.7ks exposure.
The photon index determined from the 2Ms exposure is 1.8.
This galaxy is also a radio source with a steep radio
spectrum ($\alpha$=1.0).  Its position is coincident with
an ISO source from the Aussel \etal (1999) supplementary table of
lower significance detections.
H00 note that the X-ray luminosity
for this source is consistent with that expected from hot gas
in an elliptical but the nuclear variability detected here
suggests this galaxy may also host an AGN.

{\it 3-659.1:}
The variable source is a spiral galaxy at z=0.401 (Barger \etal 2002) with
a variability significance of 4$\sigma$.  This optical galaxy is
$\sim$1.2$\arcsec$ from the hardest X-ray source ($\Gamma$=0.56) in the HDF.
The X-ray position is only 0.22$\arcsec$ from the second
reddest source in the HDF NICMOS survey (Dickinson \etal 2000).
Even though the optical galaxy is positioned further from the
center of the X-ray emission,
the detected variability suggests that perhaps some of the X-ray emission
originates from the galaxy nucleus.  There is also significant 1.4GHz
radio emission (Richards \etal 1999a) and 15$\micron$ emission detected
with ISO (Aussel \etal 1999).  The
radio position is again more closely aligned with the NICMOS source
than the optical galaxy (see Figure 1 in Richards 1999b) but both
the NICMOS and optical sources are within the positional
uncertainties.

{\it 3-965.111111:}
This is the most significant variable in our survey (9.4$\sigma$).
It was not detected in either the radio or mid-IR bands and
was first detected in the 1Ms Chandra survey
(Brandt \etal 2001b).  It was detected in both the soft and hard X-ray bands in
the 2Ms survey and has a photon index of 0.91.
It is a bright elliptical galaxy near the edge of
the HDF at z=0.663.

{\it 4-752.1:} 
This optically variable (3.1$\sigma$) X-ray source is also an FR I radio
galaxy (Richards 2000) with extensive radio structure.  The host galaxy
is a red elliptical at z=1.05.  This source has not been detected in the
hard X-ray band and thus appears to fall into the low X-ray
luminosity/soft X-ray spectrum grouping of very red
objects (Hogg \etal 2000).  Because of its faintness, the photon
index cannot be determined.  B01a suggests that the X-ray emission
is associated with the central AGN or the hot interstellar medium
and X-ray binaries of the elliptical galaxy.  It is not detected
in the mid-infrared.

{\it 4-976.1:}  
This bright spiral galaxy at z=0.089 is variable at the
3.6$\sigma$ significance level.  B01a notes that the position of
the X-ray source is not coincident with the galaxy nucleus (which is 
1.1$\arcsec$ away)
but does lie near ($\sim$0.14$\arcsec$) an off-nuclear bright spot.
They suggest that this bright spot may be a background AGN, 
starbursting region, or ``super Eddington" X-ray binary (B01a, H00).
Our variability measurements have been made on the central
nuclear region of this galaxy which, if hosting an AGN, may be 
responsible for at least part of the X-ray emission.  The photon
index for this source is 1.7.
This galaxy is within 3.3$\arcsec$ of a mid-IR source (Aussel \etal 1999). 

{\it 3-386.111:}
This is a spiral galaxy at z=0.474 which is listed in the
2Ms X-ray survey supplementary catalog containing lower significance
sources associated with optically bright objects.  The nucleus
has a variability significance of 3.8$\sigma$.
This source was not detected in the hard X-ray band.  It is
associated with an ISO source listed in the supplementary
ISO source catalog. 

One additional variable nucleus, {\it 4-254.0}, 
matched the position of a lower
significance X-ray
source indicated in the 1.38Ms Chandra survey
(Brandt \etal 2002).  This X-ray
source, however, was not listed among the 2Ms detections.
The object is an elliptical galaxy at z=0.901 which is also
associated with an ISO detection in the 
supplementary catalog.  

Of the remaining 8 variable nuclei, one
({\it 2-860.0}) is associated with a mid-IR detection and
another ({\it 3-943.0}) is within $\sim$2$\arcsec$ of
a 1.4GHz radio source.  The other 6 are not associated
with any of the multi-wavelength surveys discussed here.

Overall, 44\% (7/16) of our variable galaxy nuclei have X-ray counterparts from
the 2Ms Chandra survey (or 31\% excluding the two matches with greatest 
positional offsets).  Optical variables make up 39\% of the X-ray sources.
Of the brightest X-ray sources (full-band
flux $\geq$2.4$\times$10$^{-16}$ ergs/s) five out of nine (56\%)
are associated with optical variables.  
If we assume that the observed optical variability
indicates the presence of an AGN, this is an important check on the nature
of the X-ray emitting source and could potentially help discriminate
between low-luminosity X-ray emitting AGN and other X-ray emitting
phenomena such as supernova remnants or X-ray binaries.  
There does not appear to be
a relation between photon index and the detection of optical variability. 
Five of the eight X-ray sources in the HDF with enough signal to determine
photon indices are optical variables.  
The X-ray/optical variables make up two of the softer X-ray
sources and three of the harder sources in this small sample.

Seven of our 16 variables are associated with
mid-IR detections (44\%) and five (31\%) have radio emission at 1.4GHz.  
Variable galaxies make up 22\% of the detected sources at 15$\micron$.
This is comparable to the portion of the mid-IR integrated light
attributed to AGN based on the correlation of mid-IR sources with
Chandra sources (Elbaz \etal 1999).  

Finally, we have compared our results with two 
photometric studies to identify QSOs in the HDF on the basis of 
multi-band colors.  Jarvis \& MacAlpine (1998) identified 12 high redshift 
(z$>$3.5) QSO candidates.  Eleven of these sources
were included in our variability survey but none showed
significant variability amplitudes, having a median $\sigma$ of $\sim$0.6.
Based on structure functions for AGN/QSOs (Trevese \etal 1994; 
Hook \etal 1994; Hawkins \etal 2002) and taking into account
time-dilation effects ($\Delta$t$_{rest}$=$\Delta$t/(1+z)), the
average magnitude change expected in our survey for high-z QSOs
is $\sim$0.1--0.2 magnitudes. 
Even sources dominated by an AGN component, as expected for these
color-selected candidates, would be difficult to detect above
the 3$\sigma$ significance threshold due to their faint apparent magnitudes 
(V$_{nuc}$$\gea$27.7).  Nonetheless, an average magnitude
change of only $\sim$0.05 is significantly less than
the 0.1--0.2 average magnitude change predicted from QSO structure functions.

Conti \etal (1999) also identify 20 compact sources having QSO-like colors
and morphologies with estimated redshifts
between z$\sim$1 and 5.5.  The nuclear magnitudes for these objects
lie at V$_{nuc}$$\simeq$27--28.  
None were identified as variable above the 3$\sigma$
threshold in our survey.  The average nuclear magnitude
of this sample is $\sim$1 magnitude brighter than the Jarvis \& MacAlpine QSOs.
If these are variable QSOs/AGN-dominated galaxies, we would again expect
an average magnitude change of $\sim$0.1--0.2.
The average change in magnitude for the Conti \etal sample
over the 5 year interval is $\sim$0.03 magnitudes, with all sources lying well
below the 2$\sigma$ significance limit.

\section{OPTICAL SPECTRA}

The extensive redshift surveys of the HDF from Cohen \etal (1996; 2000),
Phillips \etal (1997) and Barger \etal (2002)
have revealed only two broad-line AGNs (BLAGNs) in the HDF proper.
These are {\it 2-251.0} 
(z=0.96 variable galaxy and X-ray source discussed above)
and {\it 4-852.12} (z=0.943 X-ray source).  The measured variability 
significance
of {\it 4-852.12} is 0.35$\sigma$.
It is possible that this source is variable but
has been observed at two points in its light curve that are close to
the same magnitude.  Further monitoring of the HDF would be necessary
to rule out optical variability for this BLAGN. 

We have studied the available spectra for all of the galaxies hosting
variable nuclei
to determine if any of the sources show
specific emission lines or line flux ratios indicative of Type 2 AGN.
Of the 16 variables, optical spectra exist for 13.
One of these is the BLAGN {\it 2-251.0} already discussed, which shows broad
MgII ($\lambda$2800) emission and absorption in its spectrum.
Almost all of the remaining 12 show emission lines in their
spectra.  Nine out of 11 show OII($\lambda$3727) when in range, with
several sources also displaying H$\beta$ and OIII($\lambda$5007).
The X-ray and strong radio FRII galaxy {\it 4-752.1}, displays
only strong absorption lines in its spectrum.  Only one source, 
{\it 4-254.0}, shows weak
NeV($\lambda$3426) emission, a line indicative of the presence of
an AGN (Hall \etal 2000).  NeIII($\lambda$3869), also stronger in
AGN than in starforming galaxies (Rola \etal 1997), is seen weakly in
2 sources, {\it 3-143.0} and {\it 3-386.111111}.  These galaxies also show
OIII and H$\beta$, but with flux ratios consistent with star formation
rather than AGN activity
(Veilleux \& Osterbrock 1997).  
Only the spiral galaxy {\it 4-976.1} is low enough redshift to
reveal strong 
H$\alpha$, SII($\lambda$6713+6731) and OI($\lambda$6300) emission in
the optical spectrum.
The line flux ratios of OIII/H$\beta$, SII/H$\alpha$ and OI/H$\alpha$
all indicate that this source is near the border that divides starforming
galaxies and AGN (Veilleux \& Osterbrock 1997).  Therefore, based on 
the optical spectra alone, objects {\it 2-251.0} (BLAGN), possibly
{\it 4-976.1} 
(LINER/Seyfert 2), and possibly {\it 4-254.0} (through the weak presence
of NeV) show evidence of AGN.

\section{LUMINOSITY FUNCTION FOR VARIABLE NUCLEI}

Knowledge of the space density of faint AGN at higher redshifts is critical 
to our understanding
of AGN evolution and the AGN phenomenon in general.  
Our sample of 16 variable sources spans a redshift range of z=0.1 to 1.8 with
most sources lying between z=0.4 and 1.1.  If we make the assumption that
these variables are indeed AGN, it is interesting to compare their
number density with local samples of AGN.  

We must first
estimate the luminosity of the AGN candidates in our survey.
Due to the faint magnitudes and small angular sizes of many
of the variables, 2-dimensional modeling to separate the components 
(disk+bulge+AGN) of the galaxies is difficult and would not yield
consistent results for all of the selected galaxies.   
A simpler approach is to estimate the
AGN luminosities by assuming that the entire nuclear flux
is attributed to the AGN.  
This value is an overestimate in one sense since the nuclear light will also
include light from the underlying galaxy, though the small nuclear aperture
minimizes the contribution from the host.  Based on the expected level
of variability over this time interval predicted from AGN structure functions,
we estimate the variable 
component for our AGN candidates to be no less than 15\% of the total nuclear
light.  In the most extreme case, the true AGN magnitude could be
$\sim$2 magnitudes fainter than that measured within the nuclear aperture.  
In another sense however, 
we underestimate the AGN flux since
we have not applied aperture corrections to include light from the wings
of the PSF.  Our aperture of r$=$0.075$\arcsec$ encircles
about 50\% of the flux from a point source (Holtzman \etal 1995).  If 
the point
source makes up the entire flux within the aperture, 
an aperture correction would increase the magnitude by up to $\sim$0.8 mag.  
If the point source is some fraction of the total nuclear light, 
the aperture correction will be smaller.  Determining the true
aperture correction is not possible without knowing 
the relative fluxes of the point source and host within 
the fixed aperture.  

Figure 8 is the absolute magnitude based on the nuclear flux
{\it vs} the source redshift (H$_o$=75 km/s/Mpc; q$_o$=0.5). 
The absolute B magnitudes
are calculated assuming a power-law index of $\alpha$=--1.0 resulting
in a K-correction of zero and V$_{F606W}$--B$_{J}$=--0.32.  
We find that most sources lie between -15$\gea$M$_B$$\gea$-17 in the
redshift range 0.4$<$z$<$1.1.  

We conduct the
Luminosity-Volume test, or V/V$_{max}$ test, of Schmidt (1968) to determine
if significant evolution is present in our sample of variable nuclei.
For this calculation, we determine for each source in our survey
the maximum redshift at which it could exist and still be included
in our survey.  The volume of space enclosed by a sphere with a
radius equal to this z$_{max}$ is the value V$_{max}$.  The 
volume enclosed at the true redshift of the source is $V$.  
Spectroscopic redshifts have been determined for 14 of our objects.  
Photometric redshifts are estimated for one other (FLY). 
There is no spectroscopic or photometric redshift published for {\it 2-456.22}.
Based on the V--I color for
this object, we estimate its photometric redshift to be z=1.8 which
corresponds to those of other HDF sources with similar
colors.  

To define z$_{max}$, we determine the limiting redshift for each source
to be detected in our survey.  The brightest limiting magnitude
in our survey
is defined by the 3$\sigma$ line in Figure 4a, which indicates the
faintest magnitude at which a source varying by a given $\Delta$m
would be considered a significant variable.  Below this magnitude,
a source varying by $\Delta$m would not be detected as variable.
Therefore, z$_{max}$ is the redshift where the apparent
magnitude of a nucleus varying by $\Delta$m falls below the 
3$\sigma$ variability significance limit shown in Figure 4a. 

The mean V/V$_{max}$ for the 16 nuclei
in our survey is 0.67$\pm$0.07.
For a population of objects uniformly distributed in space, 
$<$V/V$_{max}$$>$=0.5.
Our value of 0.67$\pm$0.07 indicates that sources are found
more often at
larger distances than at nearby distances, indicating some evolution
in the population of galaxies with variable nuclei.  
Our survey incompleteness, as discussed in Section 3, affects our
ability to detect sources
at higher redshifts and fainter magnitudes.
Thus, the value of $<$V/V$_{max}$$>$ would actually be higher if
corrected for this incompleteness.
An increase in the number density would be 
consistent with the picture of AGN/QSO evolution at higher luminosities,
where the population increases rapidly out to z$\simeq$3.

Even though evolution may be present among our sample, 
the small number of sources spread over a range in redshift makes it 
difficult to construct the luminosity function in more than one redshift bin.
To accurately compare the density of our sources to local samples,
we construct the LF in one redshift bin at z=0.4--1.1.
We choose these limits to minimize the redshift range of our sources 
while maximizing the number of sources in the bin.  
Twelve of our 16 sources lie within this redshift range, which
excludes the 2 lowest and highest redshift galaxies.  The mean
redshift for these sources is $<$z$>$=0.69. 

We calculate the luminosity function using
a technique similar in nature to the V/V$_{max}$ test described
above.  The LF is created by summing 1/V$_{max}$ for sources
in discrete magnitude bins.  In this case, we
are actually determining V$_{a}$, or the {\it accessible} volume
in which each source could be detected.  This volume is limited
not only by the magnitude limit for detection, but also by the
lower and upper redshift limits of the shell in which we are summing.
Figure 9 is the LF for HDF variable nuclei in the redshift
range z=0.4--1.1.  Incompleteness due to the fact that our variability 
selection is based on only 2 epochs is expected to produce an underestimate in
the number of AGN by a factor of $\sim$1.3, or a decrease in log$\phi$ of 
$\sim$0.12. 
For comparison, we plot the local Seyfert LFs
of Huchra \& Burg (1992; HB92) from the CfA survey and 
Ulvestad \& Ho (2001; UH01) from the Palomar survey.
Both local surveys consist of spectroscopically selected Seyfert 1s and 2s.

Our LF for variable sources at 0.4$<$z$<$1.1 extends to 
fainter absolute magnitudes than the local LFs, 
largely due to the fact that the local LFs are calculated for 
the total galaxy light (AGN+host) rather than the magnitude of the AGN alone.  
Nonetheless, many of the AGN candidates detected in
our survey may be truly fainter than those of HB92.  
Based on the spectroscopic selection
criterion employed by HB92, most of our sources would not be detected
as AGN (see Section 6) in the HB92 survey.  
Therefore, we assume that the AGN component in
our galaxies do not comprise as much of the total galaxy light as
the HB92 sources and are likely to be intrinsically fainter.  

The UH01 LF covers the same integrated galaxy magnitude range as 
HB92 but shows an overall
higher density for local Seyferts.  Ulvestad \& Ho explain that this is
due to the
fact that their LF includes intrinsically 
fainter seyfert nuclei  
than does the HB92 sample.  For the Seyfert 1 galaxies, Ho \& Peng (2001) 
decompose the nuclei from
the host galaxies in both local samples and confirm this assertion, 
finding a median nuclear magnitude of M$_B$=-14.6 for UH01 versus -17.4 for
the HB92 Seyfert 1 galaxies.  From their study, we also find that the nuclear
magnitudes of the local seyferts in both samples can be extremely
faint when compared to the total galaxy magnitude, in some cases
up to 10 magnitudes fainter.  Therefore, an LF consisting 
of nuclear
magnitudes for local Seyferts would be a much better
comparison to our HDF AGN candidates. 

Since this is not currently available,
an alternative is to produce the LF of our sources using the total
integrated galaxy magnitude rather than the nuclear magnitude alone (dashed LF).
The integrated galaxy magnitudes for our variable sources cover the same 
range as the faint end of the local Seyfert LFs but have a number density
6--10$\times$ greater than the local Seyfert density of UH01.  
If the AGN in these two samples cover a similar magnitude
range, this is evidence for a significant increase in
number density from z=0 to $\sim$0.7.  
The magnitudes for the UH01 AGN determined in Ho \& Peng (2001) for
the Seyfert 1 population cover the range -9$\gea$M$_B$$\gea$-22 mag with most
lying within -12$\gea$M$_B$$\gea$-20.  Our nuclear aperture LF (open circles
in Figure 9) extends from -14.5$\gea$M$_B$$\gea$-19.  In the most extreme
case, based on the over and underestimates of the true AGN magnitudes
discussed previously, the variable AGN in our survey could cover the 
magnitude range  
-12.5$\gea$M$_B$$\gea$-20.  This magnitude range is consistent with 
that estimated for the UH01 sample.  It is therefore unlikely that the
increase in number density between the UH01 LF and our LF is due to the 
inclusion of 
instrinsically fainter AGN as appears to be the case for the difference
between the UH01 and HB92 LFs.

Finally, we note that the LF for the variable nuclei continues
to rise at faint magnitudes, indicating that number counts have not
begun to turn over at magnitudes even as faint as M$_B$$\sim$-15.   
The effects of incompleteness in our survey due to AGN light dilution
would cause us to be less complete at the faint end and corrections
in this sense would likely increase rather than decrease our faintest
LF bin.  
Most local AGN LFs, such as those shown in Figure 9, show a flattening
toward fainter magnitudes.  The local LF of Londish \etal (2002) for
type 1 AGN reveals a much flatter slope at the faint end as compared
to the bright end and can be well fit with a two power-law function.
Our increasing numbers at the faint end may be an evolutionary
effect or might be attributed to the inclusion of a greater fraction of
obscured AGN.     
More data are needed as well as better classification into AGN types
to improve the statistics and verify this trend.

\section{SUMMARY AND CONCLUSIONS}

We have investigated variability within the Hubble Deep Field North
with the aim of identifying low-luminosity AGN in the nuclei of 
field galaxies. 
We compare photometry of the nuclei of galaxies in the initial V-band
image obtained in 1995 and a second image obtained in 2000 
and find evidence for significant variability in 16 galaxies.
Correcting for possible spurious detections and incompleteness
results in a total of $\sim$18 sources which make up 8.3\% of
galaxy nuclei to V$_{nuc}$=27.5.
The galaxies hosting variable nuclei range in redshift from 
0.09$<$z$<$1.8 with most lying between 0.4$<$z$<$1.1.  Nuclear colors
of the variable nuclei are consistent with those of non-variable nuclei.  

Seven of our variable sources are coincident (within $\sim$1.2$\arcsec$) 
with X-ray sources detected in the 2Ms Chandra exposure (Alexander \etal 2003).
We find that 39\% of the 2Ms Chandra detections in the original HDF are
significant variables.  The optical variability observed in these X-ray sources
provides important evidence to confirm the
presence of an AGN and could potentially help discriminate
between low-luminosity X-ray emitting AGN and other X-ray emitting
phenomena such as supernova remnants or X-ray binaries.
In addition, 44\% of our sources are associated with mid-IR detections
at 15$\micron$ and 31\% are detected at 1.4GHz.

Optical spectra are currently available for 13 of the 16 variable sources.
Object {\it 2-251.0} is a broad-line AGN.  Two others show some evidence
of harboring an AGN.  Object {\it 4-976.1}, a low redshift spiral, has
emission line ratios indicative of a LINER/Seyfert 2 and {\it 4-254.0} reveals 
weak NeV emission.

Based on the V/V$_{max}$ test, there is evidence for evolution within our
sample of variable nuclei in the sense that more variables are found at
higher redshifts.  We estimate the luminosity function of variable AGN
in the redshift range 0.4$<$z$<$1.1.
This LF is a lower-limit since our variability selection technique,
employing only 2 epochs for detection, could miss $\sim$20--25\% of variable
nuclei.  The nuclear magnitudes for our variables extend to 
M$_B$$\simeq$-15.0 with a median magnitude of M$_B$=-16.3.
Computing our LF with total integrated galaxy
magnitudes, we compare to the LF of Ulvestad \& Ho (2001)
for local Seyfert galaxies.
If our AGN cover a similar magnitude range as those included in their
sample, we
find evidence for an increase in the number density
of faint AGN by a factor of 6--10 from z$\sim$0 to 0.7.
The LF for our selected variable
nuclei continues to rise at faint magnitudes.  Unlike the LFs of most QSOs and
local Seyferts, we do not see signs of flattening at the faint end.

The results of this paper can be tested and confirmed with a larger
sample of variable nuclei detected at these redshifts.
A similar variability survey is being conducted for
the Groth-Westphal Survey Strip (Sarajedini \etal 2003) over 7 years
and will produce a larger sample of variables to improve
the statistical significance of any evolutionary trends.
Additional future work includes the analysis of the currently ongoing
GOODS HST Treasury program (Dickinson \etal 2003), an
ACS survey of the sky around and including 
the HDF.  The 5-epoch survey, at intervals of $\sim$45 days, will allow
for the investigation of short-term variability as well as providing
confirmation and improved completeness for the variables 
detected in this study. 

\acknowledgments
Support for proposal GO-8389.02-97A for VS and GO-8389.01-97A and 
AR-7984.01-96A for RG was provided by NASA through a grant 
from the Space Telescope Science Institute, which is operated by the 
Association of Universities for Research in Astronomy, Inc., under NASA 
contract NAS5-26555.  We gratefully
thank Rodrigo Ibata who was the PI for obtaining the second epoch
V-band image of the HDF.  A special thanks to Amy Barger and Judy Cohen 
for supplying the Keck spectra in digital form to be analyzed for these 
sources.  Thanks also to David Koo and the referee for helpful comments 
and suggestions for improving this paper.



\begin{figure}
\plotone{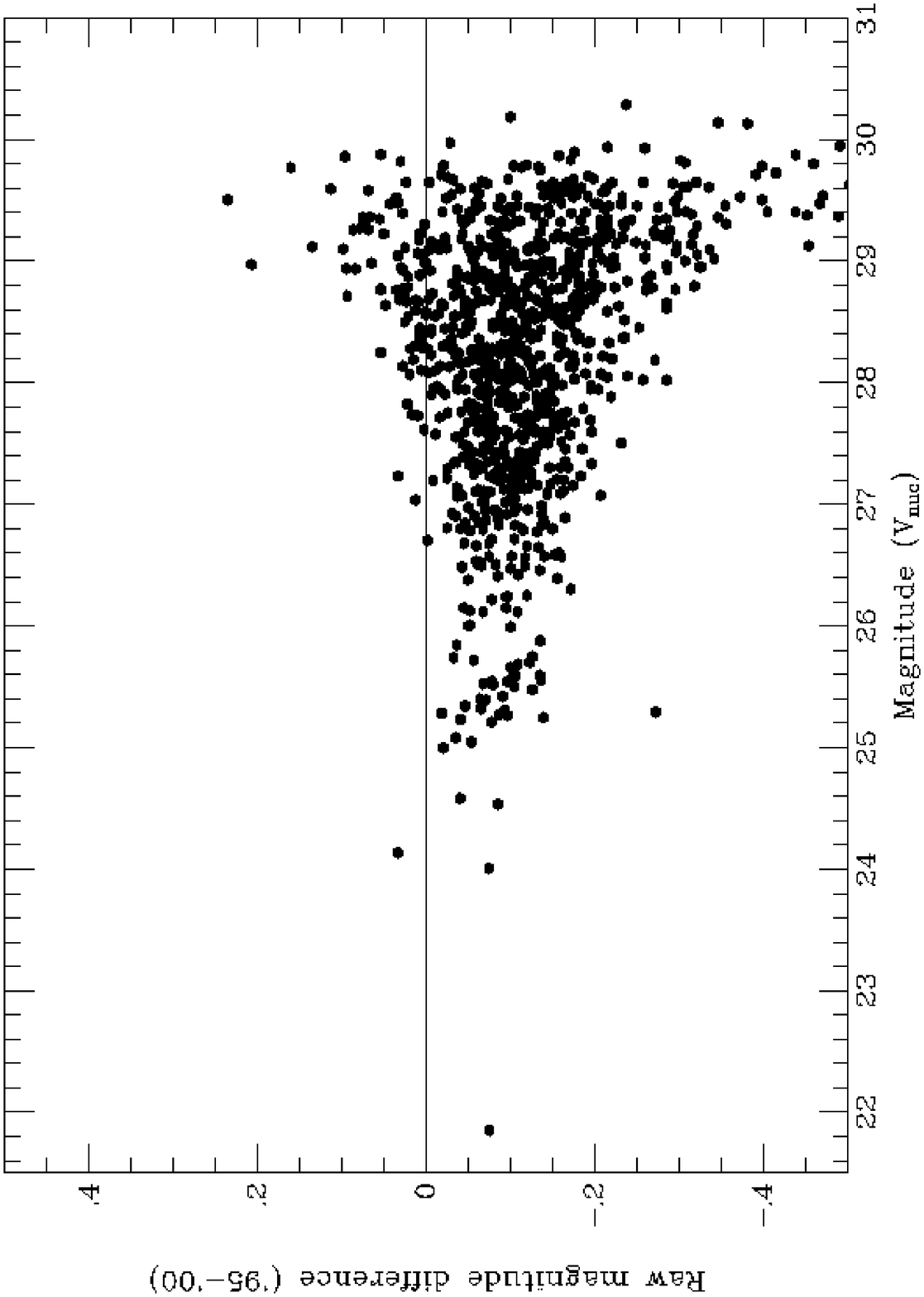}
\end{figure}

\begin{figure}
\plotone{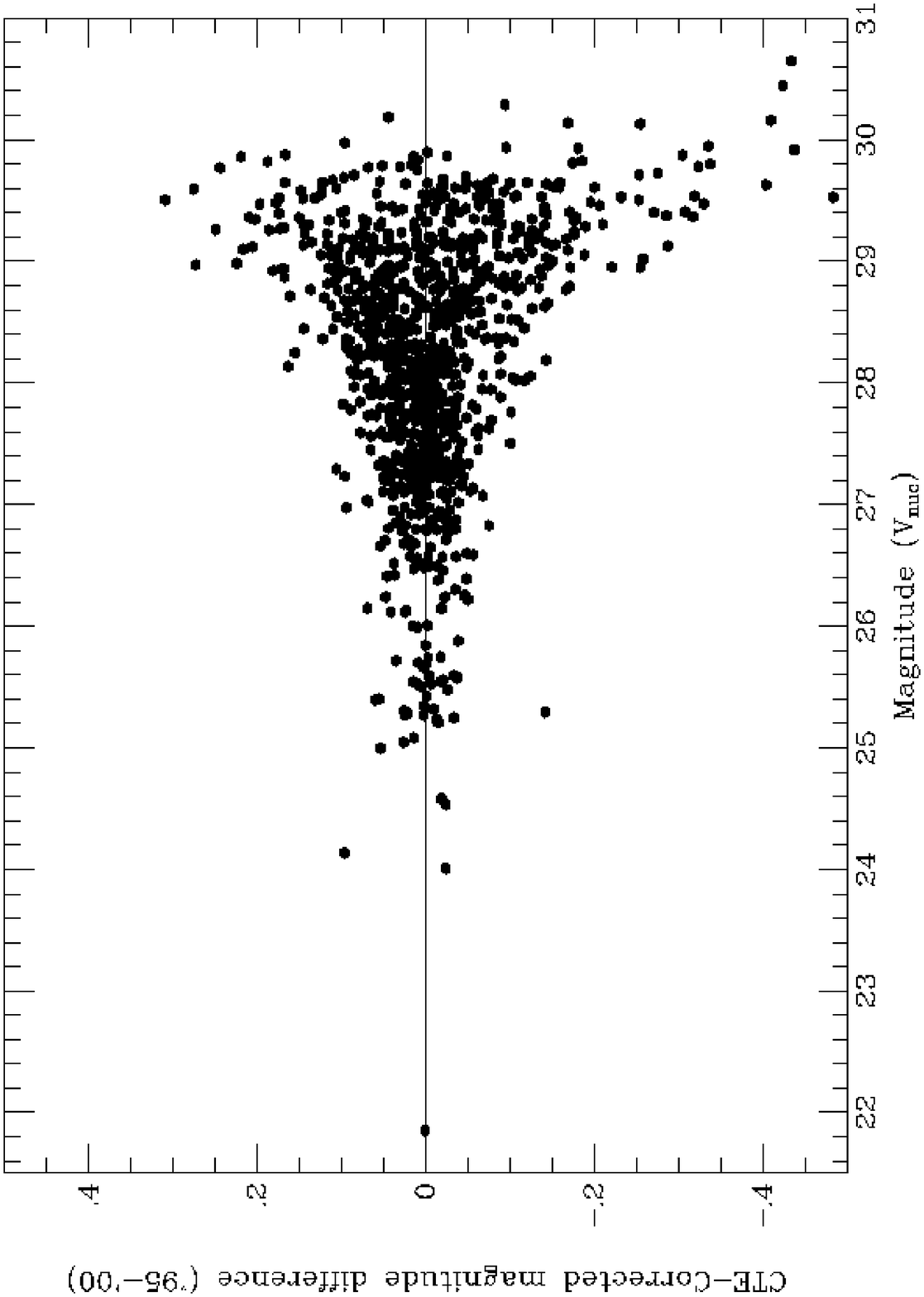}
\end{figure}

\begin{figure}
\plotone{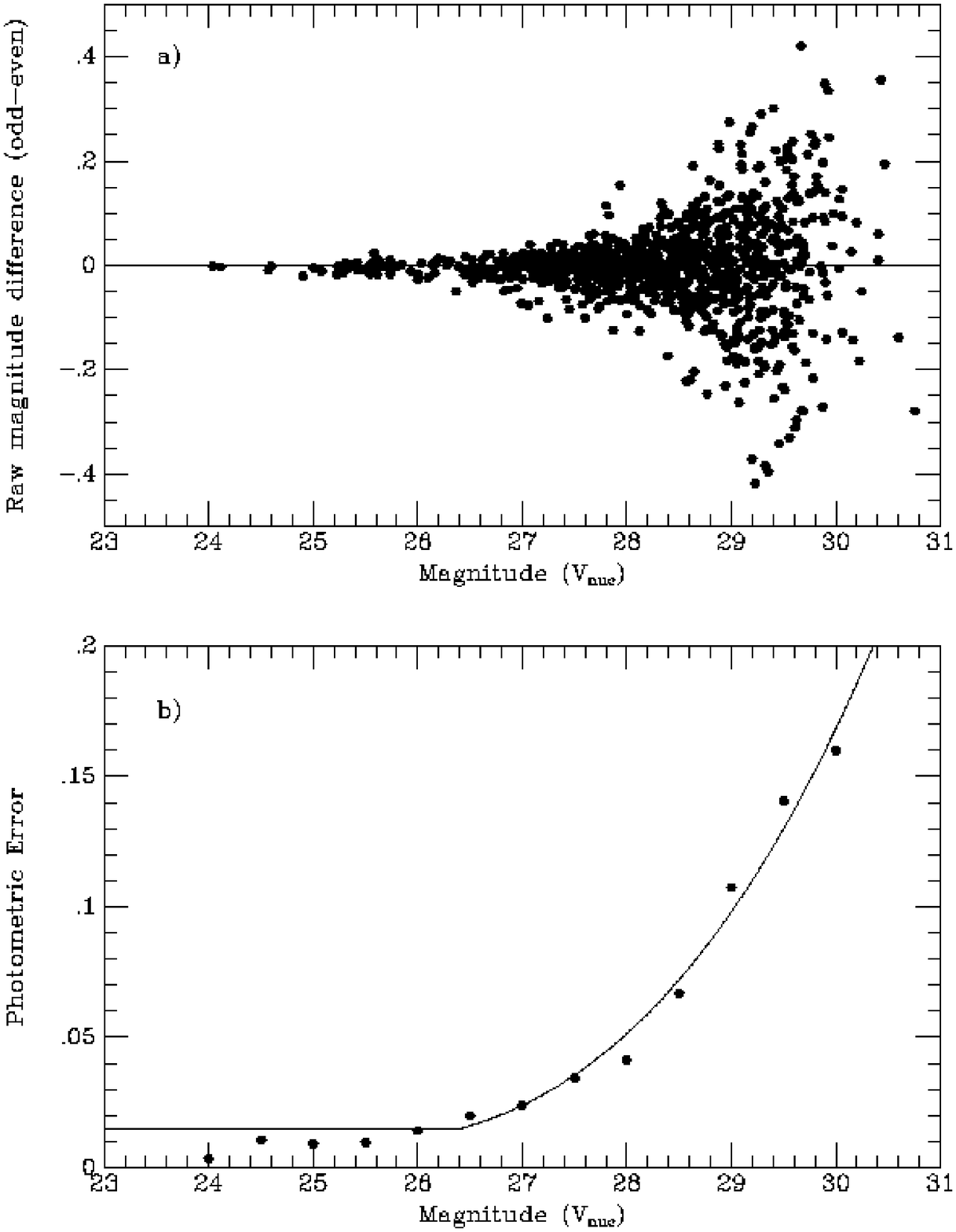}
\end{figure}

\begin{figure}
\plotone{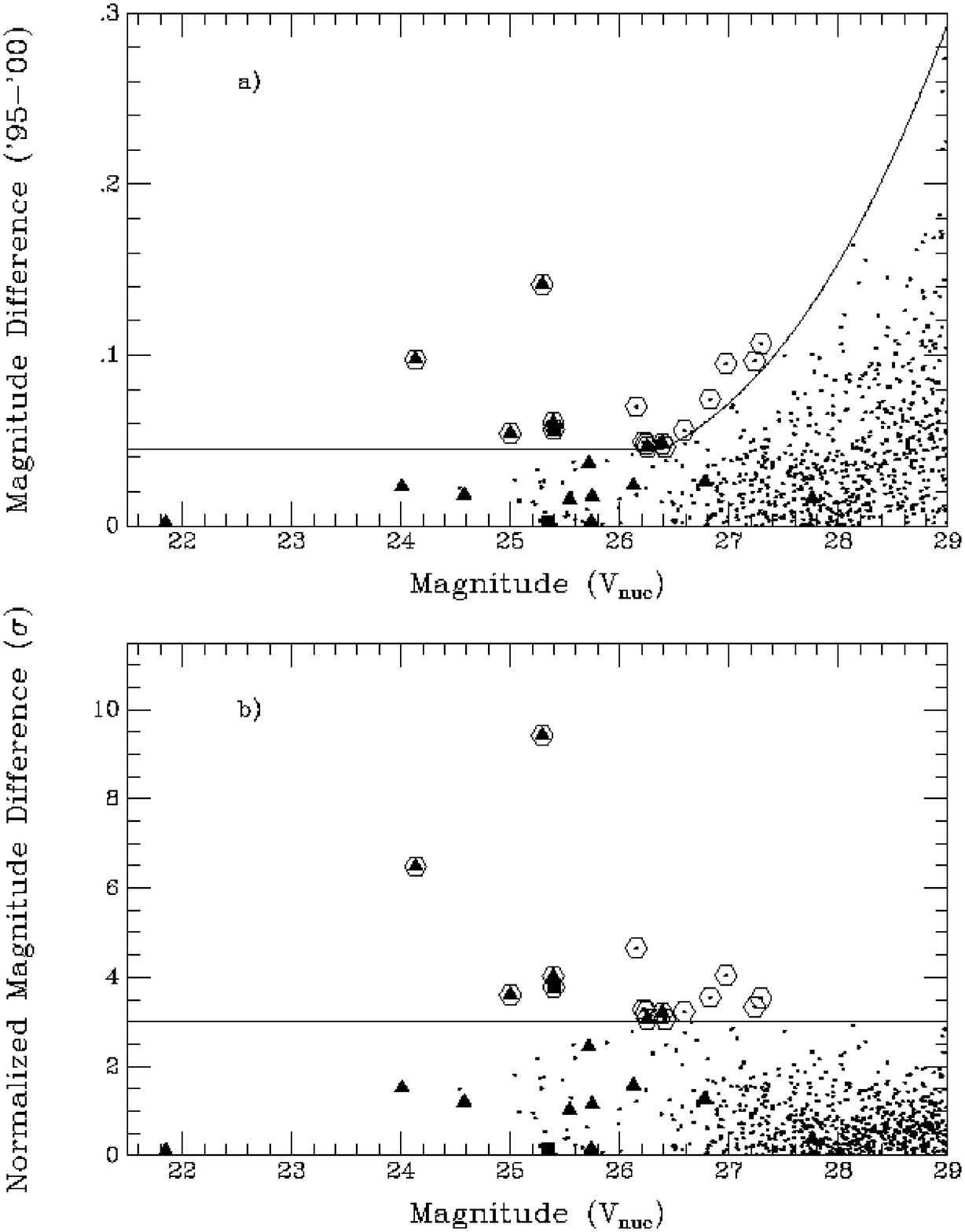}
\end{figure}

\begin{figure}
\plotone{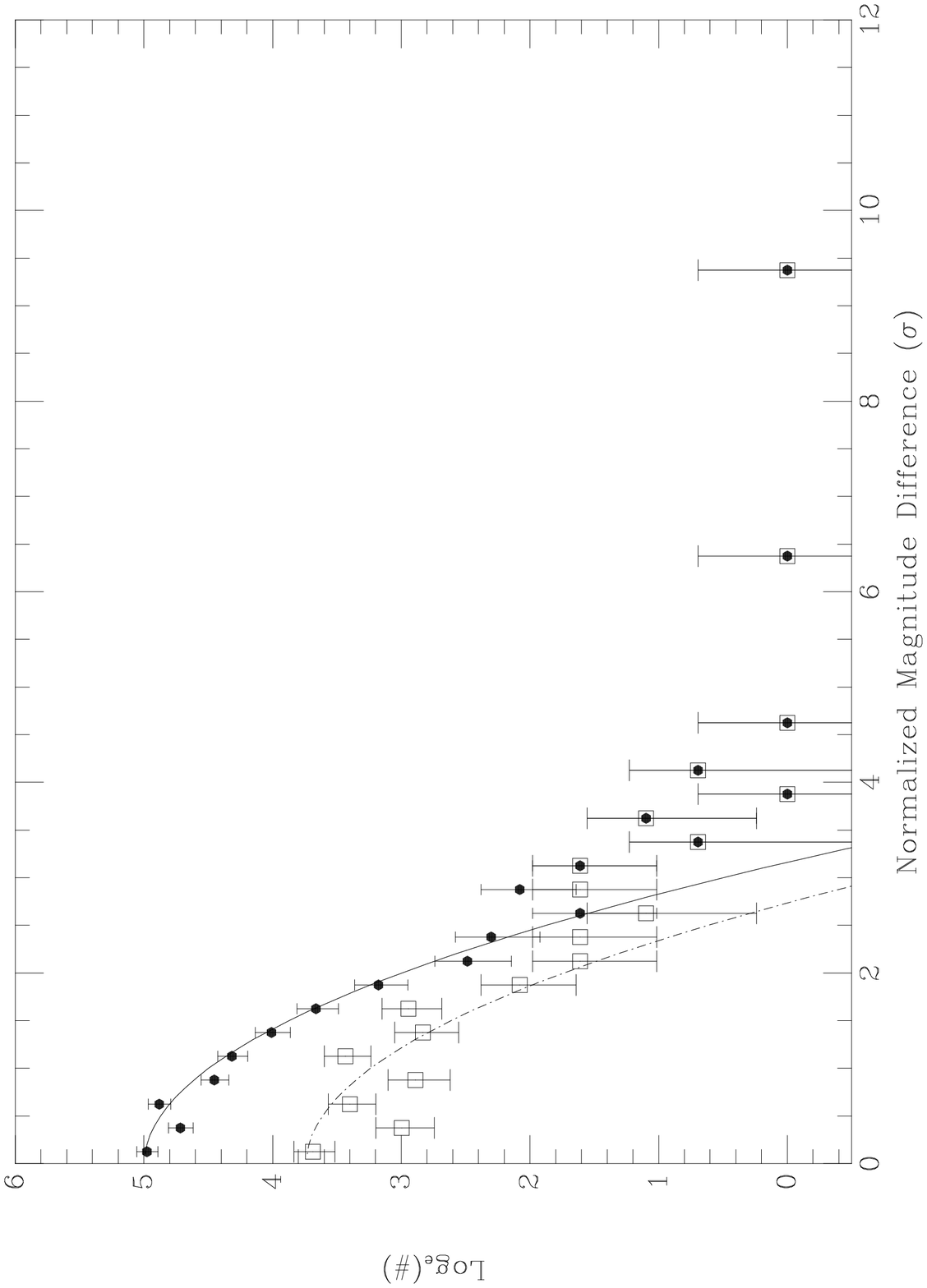}
\end{figure}

\begin{figure}
\plotone{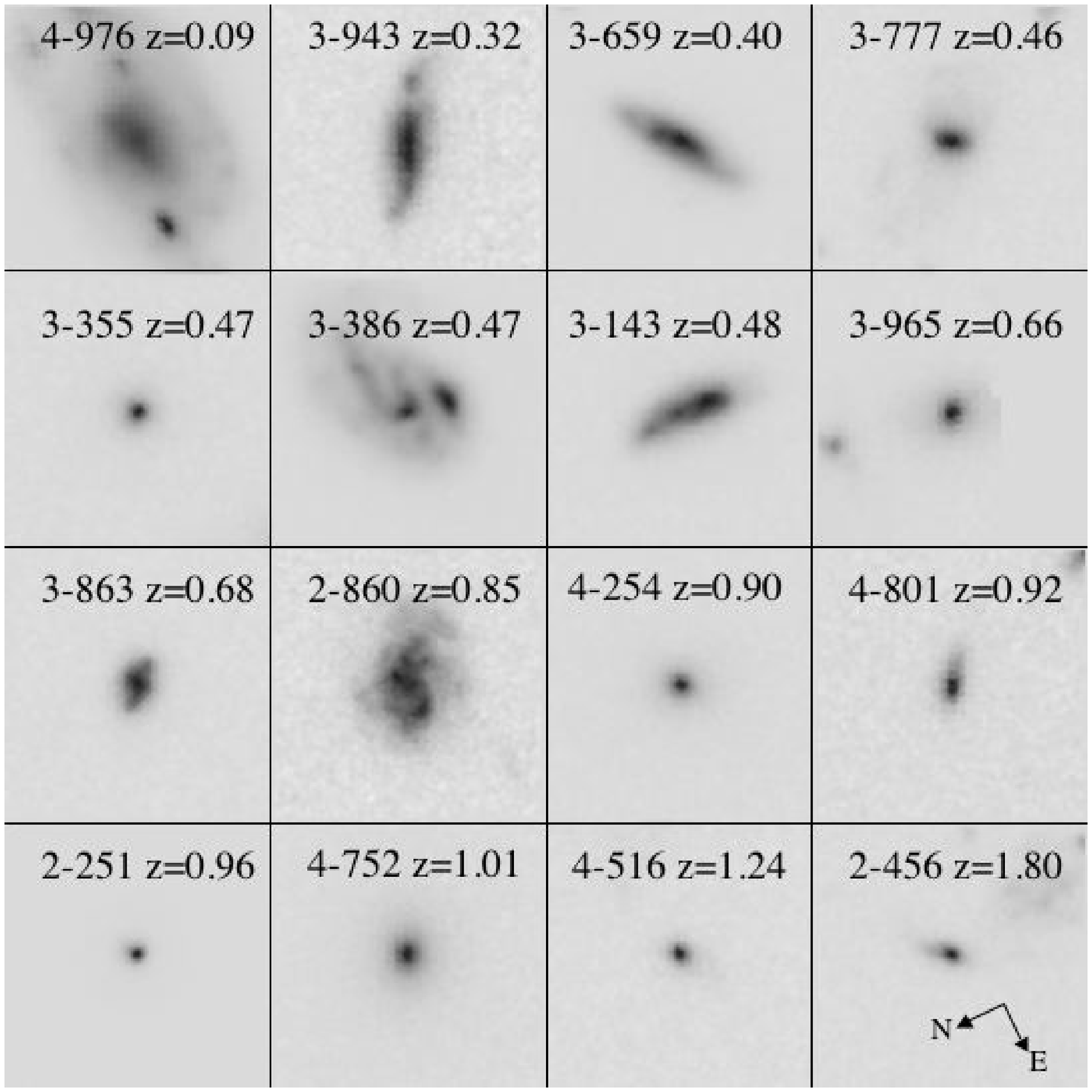}
\end{figure}

\begin{figure}
\plotone{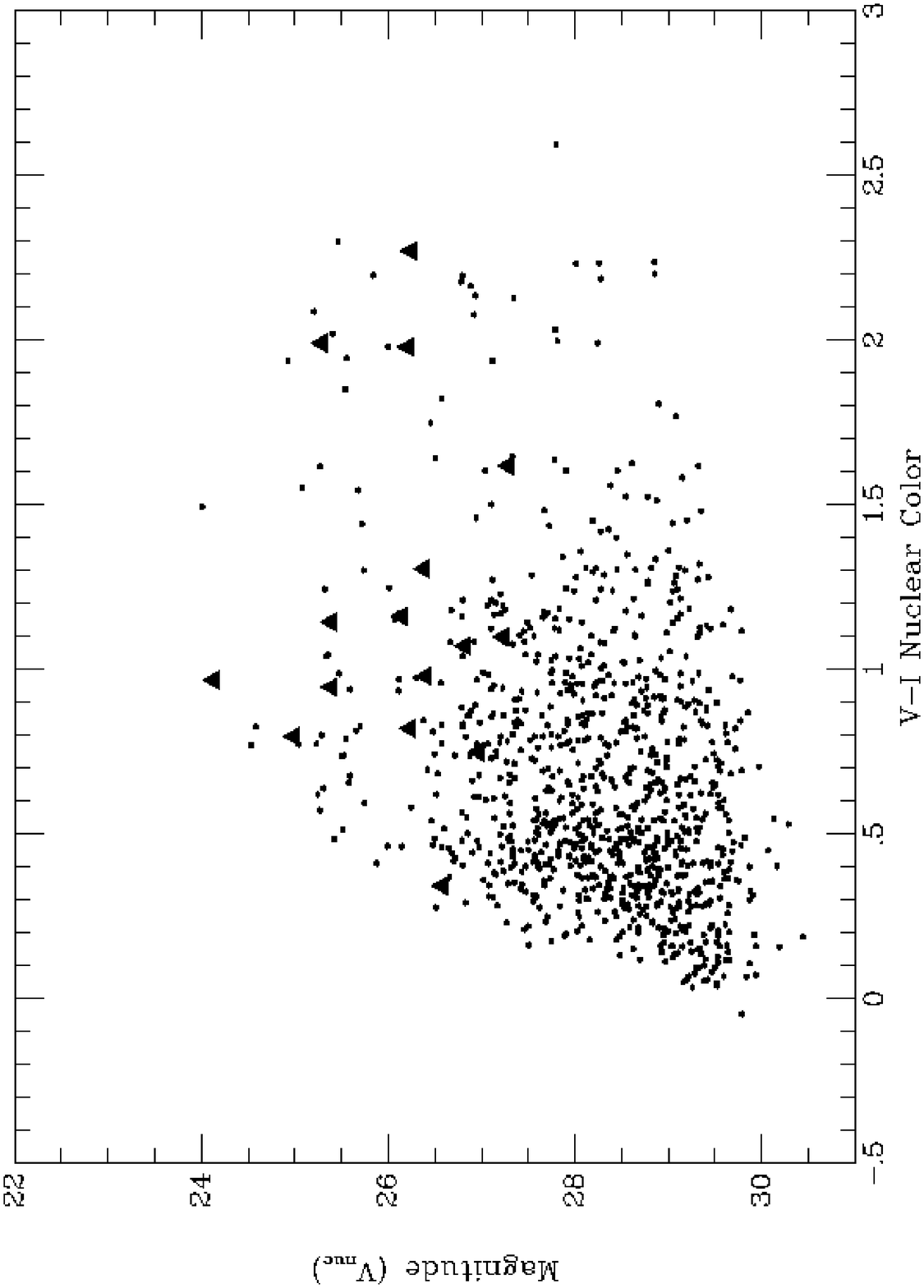}
\end{figure}

\begin{figure}
\plotone{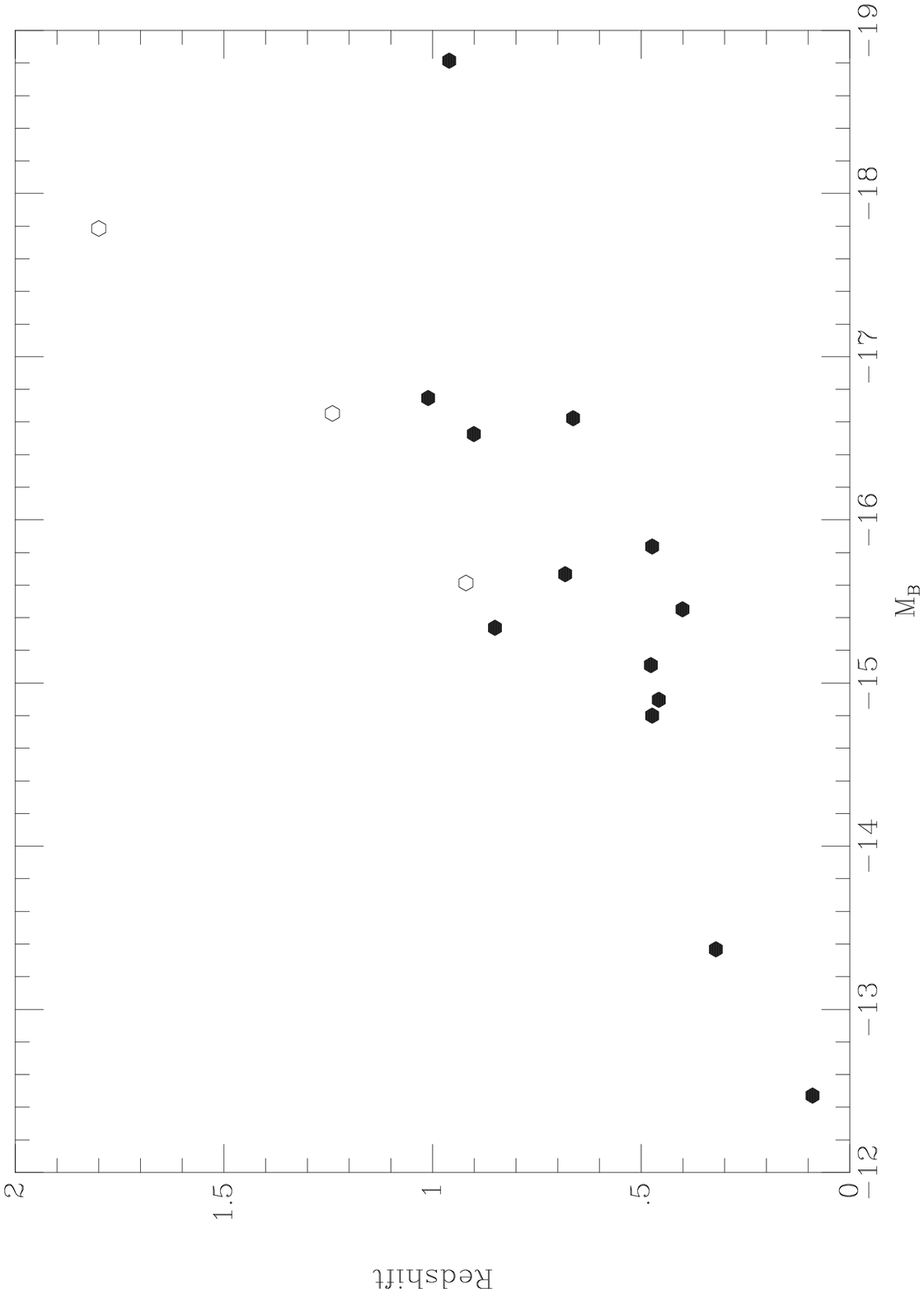}
\end{figure}

\begin{figure}
\plotone{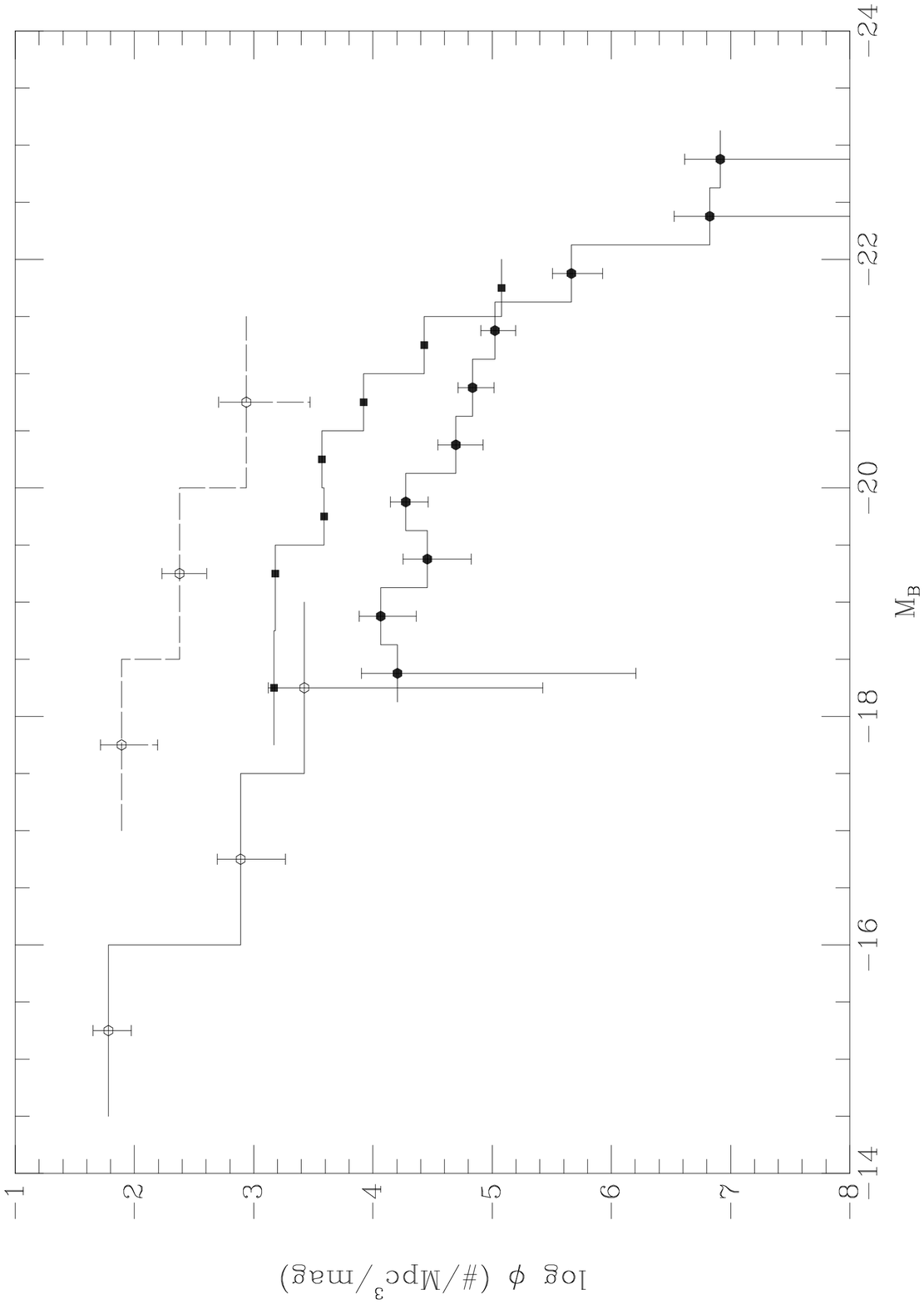}
\end{figure}


\newpage

\figcaption[f1.eps]{Magnitude difference
between the 1995 and 2000 images as a function of magnitude within an
aperture of r$=$3 pixels (0.15$\arcsec$ diameter).
The solid line represents no difference in magnitude
between the two epochs.}

\figcaption[f2.eps]{CTE-corrected magnitude difference (1995--2000).}

\figcaption[f3.eps]{
a) The magnitude difference for galaxies measured in the
odd and even data sets (see text) vs. the average magnitude
within r$=$3 pixel apertures.  b) The RMS of the magnitude differences
in unity magnitude bins.  The solid line represents a quadratic fit to the
points.}

\figcaption[f4.eps]{a) Absolute value, CTE-corrected
magnitude difference for each source in the HDF.
The solid line is the
3$\sigma$ limit indicating significant variables (3$\times$ the
solid line in Figure 3b).  Sources above this line
are marked with open hexagons.  Filled symbols indicate the
positions of Chandra X-ray sources from the 2Ms survey
(triangles are the 16 main catalog sources and squares are the
2 sources from the supplementary catalog).  b) The
magnitude differences normalized by the expected photometric error.
The Y-axis indicates the level of significance
of each source in units of $\sigma$.  Hexagons and triangles are the
same as in a).}

\figcaption[f5.eps]{Sigma distribution for galaxies
in the HDF.
The X-axis is the absolute value of the normalized magnitude difference
($\sigma$) and Y-axis is the natural logarithm of the number of points
in 0.25 mag bins.  Errorbars represent the poisson statistical errors.
Filled circles represent the 719 sources brighter
than V$_{nuc}$$=$29.0 and open squares are the 217 sources
brighter than V$_{nuc}$=27.5.
The curved lines are gaussian fits to the data within 2.5$\sigma$;
the solid line is the fit to data brighter than V$_{nuc}$=29.0 and the
dashed line is the fit to data brighter than 27.5.}

\figcaption[f6.eps]{V-band images of the 16 galaxies with variable nuclei.
Images are 3$\arcsec$ square and scaled to the same maximum pixel value.}

\figcaption[f7.eps]{
V--I Color magnitude diagram of the galaxy nuclei
in the HDF.  Filled triangles indicate the variable nuclei.}

\figcaption[f8.eps]{Absolute magnitudes of the variable nuclei
vs. redshift. Filled circles are spectroscopic
redshifts; open circles are redshifts determined indirectly
(see Table 1)}

\figcaption[f9.eps]{The luminosity function for variable nuclei in the HDF
at 0.4$<$z$<$1.1 (open circles) at $<$z$>$=0.69.
The local Seyfert luminosity
functions of Huchra \& Burg (1992; filled circles) from the CfA survey and
Ulvestad \& Ho (2001; filled squares) from the Palomar survey are shown.
The dashed LF represents the HDF variables if the total galaxy light is
included in the absolute magnitude.}


\clearpage

\begin{deluxetable}{lcclcccrccclccc}
\tabletypesize{\scriptsize}
\rotate
\setlength{\tabcolsep}{0.04in} 
\tablecaption{Variable Nuclei in the HDF}
\tablewidth{0pt}
\tablehead{
\colhead{ID\tablenotemark{a}} &
\colhead{RA (12:36)} &
\colhead{DEC (+62)} &
\colhead{Redshift} &
\colhead{Type\tablenotemark{b}} &
\colhead{B/T\tablenotemark{c}} &
\colhead{V$_{nuc}$} &
\colhead{$\Delta _{nuc}$} &
\colhead{$\sigma $} &
\colhead{X-ray ID\tablenotemark{d}} &
\colhead{offset} &
\colhead{MIR ID\tablenotemark{e}} &
\colhead{offset} &
\colhead{Radio ID\tablenotemark{f}} &
\colhead{offset}}

\startdata
2-860.0    & 54.102 & 13:54.35 & 0.851\tablenotemark{g} & 3 & 0.00 (1.25) &  27.293 &  0.107 & 3.53 & \nodata & \nodata & HDF-PM3-34 & 2.650 & \nodata & \nodata \\
2-456.22   & 50.027 & 13:51.99 & 1.8\tablenotemark{h} & \nodata & 0.52 (1.08) &  26.587 & -0.056 & 3.23 &  \nodata & \nodata & \nodata & \nodata & \nodata & \nodata \\
2-251.0    & 46.344 & 14:04.62 & 0.960\tablenotemark{i} & 2 & 1.00 (1.76)  &  24.134 &  0.097 & 6.48 &  171 & 0.194 & HDF-PM3-20 & 1.439 & 192 & 0.048 \\
3-355.0    & 56.923 & 13:01.56 & 0.474\tablenotemark{j} & 2 & 0.94 (1.00) &  26.388 & -0.048 & 3.20 & 203 & 0.066 & HDF-PS2-4 & 2.116 & 225 & 0.604 \\
3-863.0    & 58.649 & 12:21.72 & 0.682\tablenotemark{l} & 4 & 0.00 (1.14) & 26.411  & 0.046 & 3.06 & \nodata & \nodata & \nodata & \nodata & \nodata & \nodata \\
3-943.0    & 56.432 & 12:09.31 & 0.321\tablenotemark{l} & 3 & 0.00 (1.23) &  26.972 & 0.095 & 4.04 & \nodata & \nodata & \nodata & \nodata & 224 & 2.074 \\
3-143.0    & 49.644 & 12:57.43 & 0.477\tablenotemark{l} & 2 & 0.00 (1.58) &  26.148 & 0.070 & 4.65 & \nodata & \nodata & \nodata & \nodata & \nodata & \nodata \\
3-659.1    & 51.722 & 12:20.18 & 0.401\tablenotemark{j} & 2 & 0.00 (1.50)  &  25.390 & 0.060 & 4.02 &  190 & 1.235 & HDF-PM3-29 & 1.501 & 209 & 1.155 \\
3-386.111  & 50.254 & 12:39.72 & 0.474\tablenotemark{l} & 3 & 0.00 (6.23) &  25.397 & 0.057 & 3.77 &  \nodata &  \nodata & HDF-PS3-16 & 0.823 &  \nodata &  \nodata \\
3-777.1    & 52.022 & 12:09.63 & 0.458\tablenotemark{g} & 4  & 0.53 (1.14) &  26.24 &  0.048 & 3.21 & \nodata & \nodata & \nodata & \nodata & \nodata & \nodata \\
3-965.111111 & 57.485 & 12:10.55 & 0.663\tablenotemark{g} & 1 & \nodata  &  25.293 & -0.141 & 9.42 & 206 & 0.456 & \nodata & \nodata &  \nodata &  \nodata \\
4-254.0    & 46.127 & 12:46.50 & 0.901\tablenotemark{g}  & 1 & 0.90 (1.03) &  26.216 & -0.049 & 3.28 &  \nodata &  \nodata & HDF-PS3-17 & 2.182 & \nodata &  \nodata \\
4-516.0    & 45.652 & 11:53.97 & 1.24\tablenotemark{m} & 3 & 0.66 (1.08)&  26.827 & -0.074 & 3.54 &  \nodata & \nodata & \nodata & \nodata & \nodata & \nodata \\
4-752.1    & 44.377 & 11:33.20 & 1.011\tablenotemark{j}  & 1 & 0.87 (1.25)&  26.250 & -0.046 & 3.07 &  165 & 0.210 & \nodata & \nodata & 187 & 0.121 \\
4-801.0    & 39.990 & 12:33.65 & 0.920\tablenotemark{k} & 4 & 0.17 (1.07)&  27.234 & 0.096 & 3.33 & \nodata & \nodata & \nodata & \nodata & \nodata & \nodata \\
4-976.1    & 41.643 & 11:31.85 & 0.089\tablenotemark{l} &  4 & 0.00 (19.1)&  24.998 & 0.054 & 3.60 &  160 & 1.112 & HDF-PM3-12 & 3.308 & \nodata & \nodata \\ 

\enddata

\tablenotetext{a}{Galaxy ID from Williams \etal 1996.}
\tablenotetext{b}{Spectroscopic type based on photometry from FLY:
1=E/S0, 2=Sbc, 3=Scd, 4=Irr}
\tablenotetext{c}{Bulge fraction from Marleau \& Simard (1998) followed by
the chi-square goodness of fit in parantheses.}
\tablenotetext{d}{X-ray ID number from in Brandt \etal (2001).}
\tablenotetext{e}{15$\micron$ ID from in Aussel \etal (1999).}
\tablenotetext{f}{1.4GHz ID number from in Richards \etal (1999).}
\tablenotetext{g}{Redshift from Cohen \etal (2000).}
\tablenotetext{h}{FLY does not provide a photometric redshift for this source.
The authors have estimated the redshift based on the photo-z's of
other galaxies having similar colors in the FLY catalog.}
\tablenotetext{i}{Redshift from Phillips \etal (1997).}
\tablenotetext{j}{Redshift from Barger \etal (2002).}
\tablenotetext{k}{Redshift Photometric redshift from FLY.}
\tablenotetext{l}{Redshift from Cohen \etal (1996).}
\tablenotetext{m}{Redshift based on absorption feature observed in
a nearby galaxy spectrum (Bunker \etal 2000)}

\end{deluxetable}

\end{document}